# The X-Ray Telescope of the CAST Experiment

R. Kotthaus, H. Bräuninger, P. Friedrich, R. Hartmann, D. Kang, M. Kuster, G. Lutz, and L.Strüder

*Abstract*–The CERN Axion Solar Telescope (CAST) searches for solar axions employing a 9 Tesla superconducting dipole magnet equipped with 3 independent detection systems for X-rays from axion-photon conversions inside the 10 m long magnetic field. Results of the first 6 months of data taking in 2003 imply a 95 % CL upper limit on the axion-photon coupling constant of $g_{a\gamma} < 1.16 \times 10^{-10}$ GeV$^{-1}$ for axion masses $m_a < 0.02$ eV. The most sensitive detector of CAST is a X-ray telescope consisting of a Wolter I type mirror system and a fully depleted pn-CCD as focal plane detector. Exploiting the full potential of background suppression by focussing X-rays emerging from the magnet bore, the axion sensitivity obtained with telescope data taken in 2004, for the first time in a controlled laboratory experiment, will supersede axion constraints derived from stellar energy loss arguments.

## I. Introduction

THE CERN Axion Solar Telescope (CAST) experiment searches for solar axions employing a magnetic "helioscope" [1]. The existence of the axion, a pseudoscalar boson, would be the consequence of the proposed solution [2] of the "strong CP problem" caused by the apparent smallness of the electric dipole moment of the neutron. Light axions (of mass $m_a < 1$ eV) would also be viable candidates for relic cold and hot [3] Dark Matter in the Universe.

The helioscope technique to search for the hypothetical axion exploits its coupling to two photons which would lead to axion-photon conversions in external electric or magnetic fields *(Primakoff effect)*. Therefore, stars could produce axions (and generically other particles with 2-photon interactions, like gravitons) by transforming thermal photons in the fluctuating electromagnetic fields of the stellar plasma. For a terrestrial axion search the sun, of course, would be the strongest source of axions with a thermal energy spectrum peaking at about 3 keV. Fig. 1 shows the differential solar axion flux calculated [4] for the standard solar models of 1982 [5] and 2004 [6]. Solar axions arriving here on earth unattenuated could be reconverted into X-rays in a strong transverse magnetic field B with a conversion probability $P_{a\gamma} = g_{a\gamma}^2 (B/q)^2 \sin^2(qL/2)$, where $g_{a\gamma}$ is the axion-photon coupling strength, $q = m_a^2/2E$ the axion-photon momentum difference, E the axion energy and L the length of the magnetic conversion region. Within the limit of full axion-photon coherence ($qL \ll 1$) $P_{a\gamma} = g_{a\gamma}^2 (BL)^2 / 4$. Otherwise $P_{a\gamma}$ is reduced due to the axion-photon momentum mismatch.

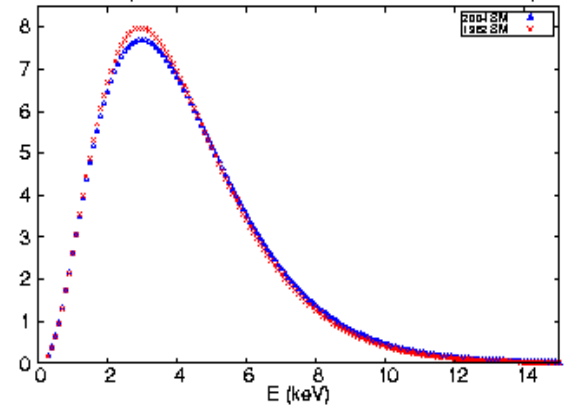

Fig. 1. Solar axion flux calculated [4] for a modern solar model [6] (2004 SM, blue triangles) and for an older one [5] (1982 SM, red crosses). The flux is given in units of $10^{10}$ axions cm$^{-2}$ sec$^{-1}$ keV$^{-1}$ for an axion-photon coupling strength of $g_{a\gamma} = 10^{-10}$ GeV$^{-1}$.

The CAST helioscope consists of a superconducting dipole magnet originally developed at CERN as a prototype bending magnet for the *Large Hadron Collider (LHC)* and 3 independent detection systems for X-rays in the energy range of about 1 to 10 keV. The magnet produces a homogenous transverse magnetic field of 9 Tesla inside two 9.26 m long tubes of 43 mm diameter each. A rotational support allows to point the magnet to the sun for about 1.5 hours twice a day around sunrise and sunset. The most sensitive X-ray detector of CAST is a telescope (Fig. 2) employing a Wolter I type grazing incidence X-ray mirror system and a fully depleted pn-CCD as focal plane detector. Both instruments were available from X-ray satellite missions.

Manuscript received November 11, 2005.

This work was supported in part by the German Bundesministerium für Bildung und Forschung (BMBF) under Grant No. 05CC2EEA/9, in part by the N3 Dark Matter network of the Integrated Large Infrastructure for Astroparticle Science (ILIAS), and the Virtuelles Institut für Dunkle Materie und Neutrinos (VIDMAN).

R. Kotthaus is with the Max-Planck-Institut für Physik, Werner-Heisenberg-Institut, D-80805 München, Germany (telephone: (89)-32354-265, e-mail: rik@mppmu.mpg.de).

H. Bräuninger is with the Max-Planck-Institut für extraterrestrische Physik, D-85748 Garching, Germany (telephone: (89)-745577-11, e-mail: hb@mpe.mpg.de).

P. Friedrich is with the Max-Planck-Institut für extraterrestrische Physik, D-85748 Garching, Germany (telephone: (89)-30000-3515, e-mail: pfriedrich@mpe.mpg.de).

R. Hartmann is with PNSensor GmbH, D-80803 München (telephone: (89)-839400-46, e-mail: roh@hll.mpg.de).

D. Kang is with the Physikalisches Institut, Albert-Ludwigs-Universität Freiburg, D-79104 Freiburg, Germany (telephone: (761)-203-5877, e-mail: donghwa@physik.uni-freiburg.de).

M. Kuster was with the Max-Planck-Institut für extraterrestrische Physik, D-85748 Garching, Germany. He is now with the Institut für Kernphysik, Technische Universität Darmstadt, D-64289 Darmstadt, Germany (telephone: (6151)-16-2321, e-mail: kuster@hll.mpg.de).

G. Lutz is with the Max-Planck-Institut für Physik, Werner-Heisenberg-Institut, D-80805 München, Germany (telephone: (89)-839400-21, e-mail: gel@hll.mpg.de).

L. Strüder is with the Max-Planck-Institut für extraterrestrische Physik, D-85748 Garching, Germany (telephone: (89)-839400-41, e-mail: lts@hll.mpg.de).

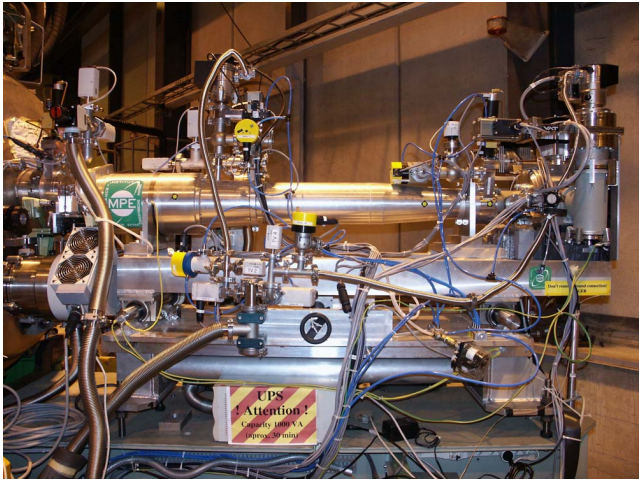

Fig. 2. X-ray telescope of the CAST experiment. The telescope is attached to one of the 2 magnet bores (outside the picture to the left) at the end facing the rising sun. The X-ray mirror is mounted inside the left-hand section of the telescope. The focal plane containing the pn-CCD detector is at the right-hand end.

The telescope focuses X-rays from solar axion conversions inside the magnet bore of 14.5 cm$^2$ aperture to a spot size of about 6 mm$^2$, thus reducing the sensitive area of the focal plane detector required to accept the full axion signal from the solar core (inner 20 % of radius) by a factor of more than 200 and enhancing the signal-to-background ratio correspondingly. Moreover, the concentration of the axion signal to a small fraction of the total sensitive area of the pn-CCD leaves the remainder of the detector for simultaneous background measurements thus eliminating systematical uncertainties due to background subtraction from sequential measurements.

The combined use of instrumentation developed for high energy physics and for astrophysics enhances the solar axion sensitivity ($\sim g_{a\gamma}^4$) of CAST by more than 3 orders of magnitude compared to the previously most sensitive helioscope [7].

CAST has been taking data since May of 2003 and first results of a 6-months run in 2003 have been published [8]. No solar axion signal was observed and the corresponding 95 % confidence level (CL) upper limit on $g_{a\gamma}$ is shown in Fig. 3 together with results from previous searches, with the best limit inferred from astrophysical arguments and with axion model predictions. Since then, the CAST telescope has been upgraded and from May to November 2004 has performed 115 tracking cycles at sun rise for a total of 179.4 hours. In addition, background data were taken for 1723.5 hours with the telescope not pointing at the sun.

In the following we will describe the X-ray telescope, present the performance in the 2004 data taking and discuss future prospects. For previous accounts of the CAST X-ray telescope we refer to [9] - [11].

## II. THE X-RAY TELESCOPE

THE CAST X-ray telescope (Fig. 2) consists of components developed and built for X-ray astrophysics. The X-ray mirror system is a spare flight module of the satellite mission *ABRIXAS* [12]. The focal plane detector is a pn-CCD of the type being used since more than 5 years in *XMM-NEWTON* [13]. The properties and performance of the 2 telescope components are more than adequate for a solar axion search.

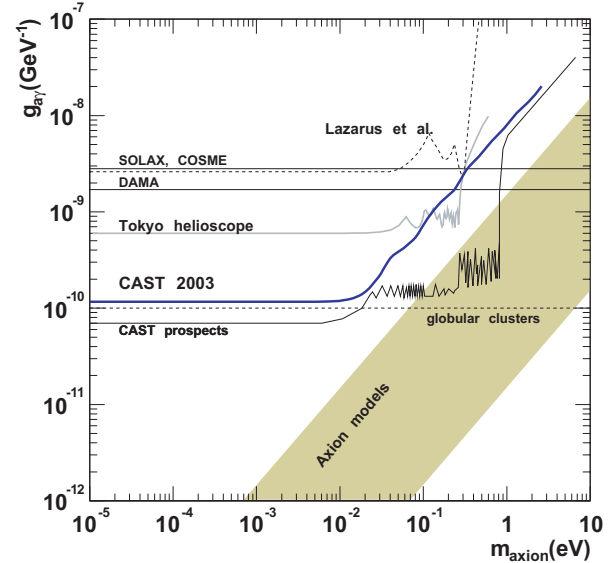

Fig. 3. Exclusion limit (95 % CL) from the CAST 2003 data [8] and the expected sensitivity of forthcoming CAST results compared with previous limits, with an astrophysical constraint ("globular clusters") and with axion model predictions.

### A. The X-ray mirror system

The Wolter I type X-ray mirror system (Fig. 4) of 1600 mm focal length consists of 27 Au-coated parabolic and hyperbolic mirror shells nested in a spoke structure subdividing the mirror aperture into 6 azimuthal sectors.

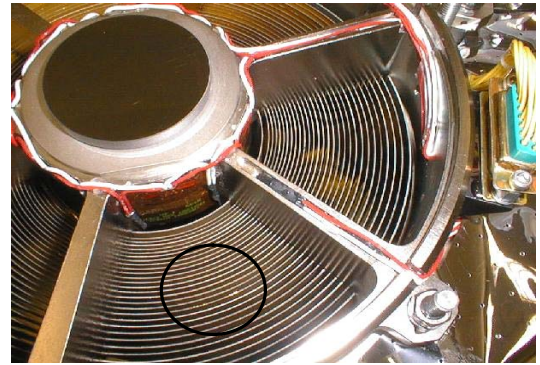

Fig. 4. Front view of the mirror system of the CAST telescope. 27 concentric mirror shells are supported by a spoke structure subdividing the aperture into 6 azimuthal sectors. The center of one of these sectors is used to focus x-rays entering the mirror from the magnet bore (approximate projected size and location indicated by black circle).

Since the diameter of the outermost mirror shell (163 mm) is much larger than the opening of the CAST magnet the mirror is mounted off-axis such that only the central part of one mirror sector is being used for imaging and acceptance losses are minimized. The overall X-ray detection efficiency given by the mirror reflectivity and the quantum efficiency of the pn-CCD is shown in Fig. 5. The mirror reflectivity depends on the mounting of the telescope and the alignment of the telescope and magnet bore axes. The curves in Fig. 5 are the result of ray tracing simulations which were compared to transmission measurements at the *PANTER* test facility in Munich prior to the installation of the mirror at CAST [9]. The integral detection efficiency of the telescope for X-rays from axion conversions is 30 to 35%. The high angular resolution of the telescope of about 40 arcsec half energy width [9], which is needed for the observation of astrophysical point sources, the application the mirror was designed for, is by far not exploited in CAST. The divergence of the axion beam from the core of the sun would be larger by a factor of 10.

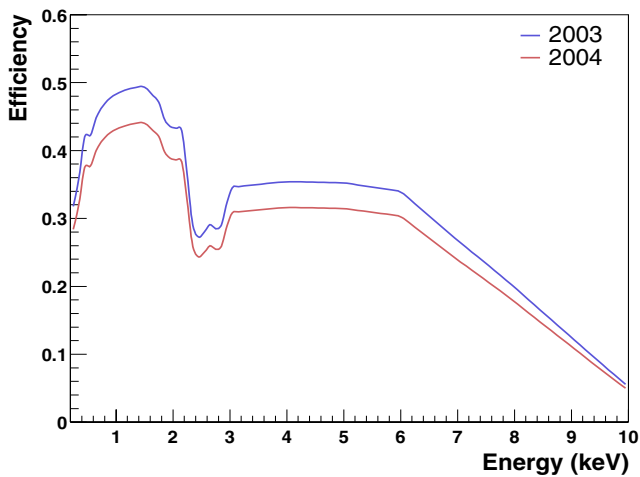

Fig. 5. Overall X-ray detection efficiency of the CAST telescope given by the mirror reflectivity and the quantum efficiency of the pn-CCD. For the data taking period in 2004 the efficiency is reduced by 11% compared to 2003 due to a small adjustment of the telescope axis necessary for a better centering of the solar axion spot on the CCD sensitive area. The reduced efficiency above 2.2 keV is due to absorption effects at the M-edge of the Au coating of the mirror surface

### B. The pn-CCD Detector

The focal plane detector (Fig. 6) of the CAST telescope is a 280 μm thick fully depleted pn-CCD with integrated front-end electronics [13]. The sensitive area of 0.96 cm x 3 cm is subdivided into 64 x 200 pixels of 150 μm x 150 μm. For the telescope focal length of 1600 mm the pixel size corresponds to a projected angular acceptance of 19.3 arcsec. The "axion image" of the core of the sun would have a diameter of 19 pixels or slightly less than 3 mm. The detector is operated at −130°C and housed inside a shield (removed in Fig. 6) of typically 10 mm of Cu and more than 20 mm of Pb. The 64 columns of 200 pixels are read out in parallel. The readout time is 6.1 msec per cycle of 71.8 msec.

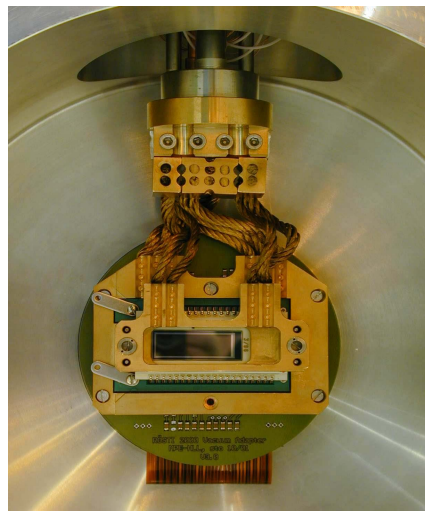

Fig. 6. Focal plane pn-CCD detector as mounted inside the CAST telescope. The Au-plated cooling mask surrounding the rectangular CCD chip in the center is connected to the cold finger of a Stirling cooler on top of the detector chamber. Electrical connections to the printed circuit board behind the CCD chip are provided via the flexlead leaving the chamber at the bottom. The internal Cu/Pb shield surrounding the detector has been removed.

The most important property of the pn-CCD for the use in CAST, next to the spatial resolution, is the high and smooth quantum efficiency (QE) in the entire photon energy region of interest for the solar axion search. Fig. 7 shows the QE measured for the device used in *XMM-NEWTON* [13]. In the range from 1 to 7 keV, which contains more than 90 % of the axion signal (see Fig. 1), the QE exceeds 95%.

Another valuable feature of the pn-CCD with integrated frontend readout electronics is the excellent longterm stability of operating parameters and performance resulting in homogenous data sets collected over longer periods of time.

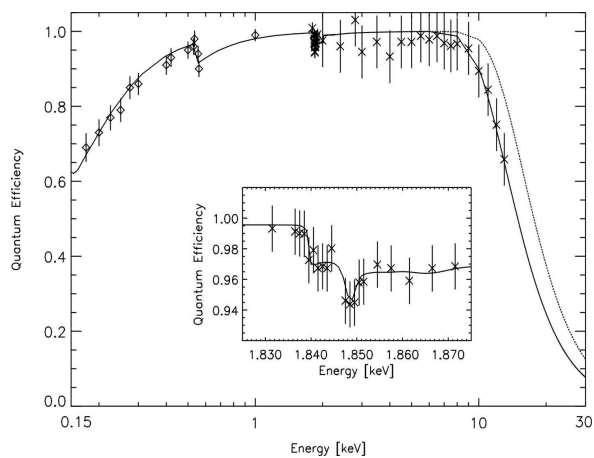

Fig. 7. Quantum efficiency (QE) of the fully depleted pn-CCD as measured for the *XMM-NEWTON* device [13]. The drop of QE at 0.53 keV is caused by absorption losses in the $SiO_2$ passivation layer at the detector surface. The inset shows the absorption fine structure of the Si K edge at 1.84 keV. The solid line is a detector model fit to the measurements. The dotted line would be the QE of a 500 μm thick detector.

The performance of the CAST pn-CCD is monitored in daily flat field illuminations with a Fe$^{55}$ source. The result of these calibration measurements for the data taking period in 2004 is given in Fig. 8. The most important qualities (energy calibration, charge transfer inefficiency (CTI), uniformity of spatial response) are stable over the entire period spanning more than 6 months. The noise conditions and correspondingly the energy resolution show some variations and do not match the quality obtained under controlled laboratory conditions. We assign this to the uncontrolled and sometimes high noise level in the CAST experimental hall. In no way did the observed degradations affect the result of the axion search.

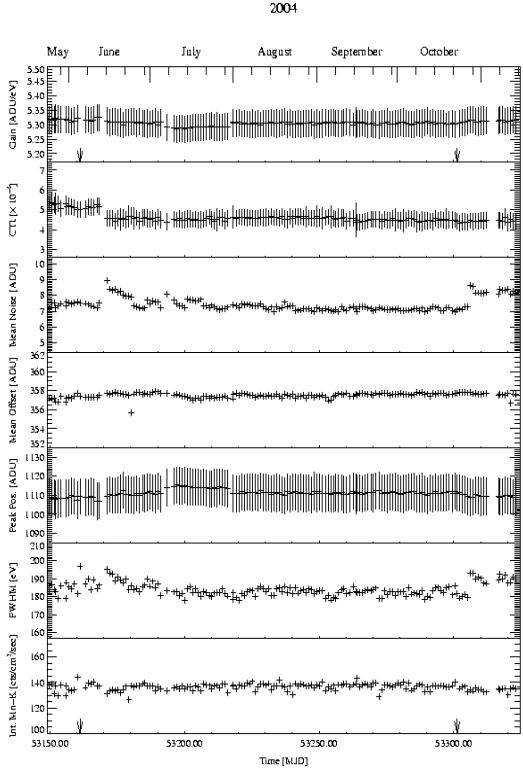

Fig. 8. Performance of the pn-CCD of the CAST telescope during the data taking period of 2004. The results are from daily calibration measurements using a Fe$^{55}$ source. From top to bottom are shown: energy calibration, charge transfer inefficiency (CTI), mean noise and offset averaged over all 12800 pixels, peak position, line width (FWHM) and intensity of Mn-K$_\alpha$ photons.

*C. Telescope Alignment and Pointing Accuracy*

Prior to the two data taking periods in 2003 and 2004 the telescope was aligned with the magnet axis using a laser beam through the entire system. The location and the size of the CCD image of a parallel laser beam filling the entire magnet bore is used to determine the solar axion spot on the CCD chip. The observed laser spot size is well within the anticipated solar axion image of 6.4 mm$^2$. In 2004, the stability of the focus during the course of the measurements was monitored by a 100 Megabecquerel pyroelectric X-ray source at the far end of the magnet. The focus of the slightly divergent beam of mainly 8 keV photons through the magnet bore is 30 cm beyond the CCD. Therefore, the X-ray image is displaced and somewhat larger than the focal spot of the parallel laser beam, but it can be used to monitor the temporal stability of the X-ray optics. The observed X-ray spot positions at the beginning and towards the end of the data taking period in 2004 and for different magnet orientations reproduced within less than one pixel (< 20 arcsec).

The overall pointing accuracy of the CAST helioscope inferred from redundant angular encoder systems and direct optical observations of the sun [14] is better than about 1 arcmin, which is perfectly adequate, given the angular acceptance of the magnet bore of 16 arcmin.

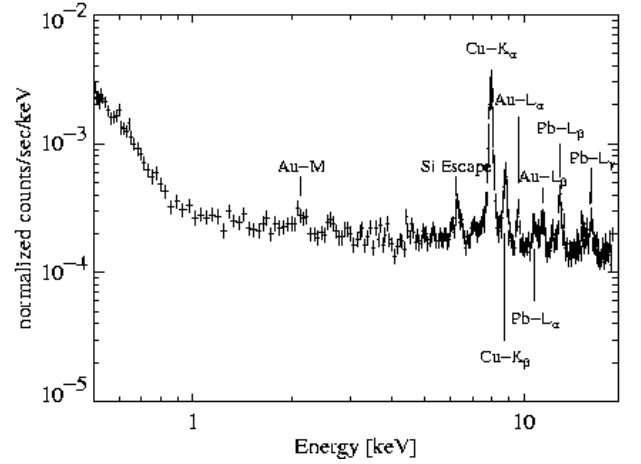

Fig. 9. Background energy spectrum of events of photon signature (single, double, triple and quadruple patterns of contiguous pixels) as measured in the entire sensitive CCD area in 1723.5 h of data taking in 2004.

III. EXPERIMENTAL RESULTS

Fig. 9 gives the energy spectrum of CCD events with valid photon signatures as measured in 1723.5 hours of data taking in 2004 with the helioscope not pointing at the sun. The dominant features of this background spectrum are due to fluorescence from Cu (K-lines and Si escape peak), Au and Pb contained in materials close to the CCD chip which had not been selected for low radioactivity. Samples of the structural materials of the detector have been probed for their contents of natural radioactivity to be put into ongoing GEANT4 model simulations [11]. The steep rise of the spectrum in Fig. 9 at energies below 0.8 keV is due to pickup of intermittent external noise produced in the CAST environment. The inherent CCD detector noise is limited to energies below about 0.3 keV. In the axion search region between 1 and 7 keV the background spectrum shows a broad minimum with little structure at an average flux level of $(7.69 \pm 0.07) \times 10^{-5}$ events cm$^{-2}$ sec$^{-1}$ keV$^{-1}$. This corresponds to an integral background count rate of 0.1 event per hour in the solar axion spot area of the CCD.

The integral background count rate in 2004 is lower by a factor of 15 compared to 2003 enhancing the sensitivity to detect an axion signal by a factor of 3.9. This is achieved by restricting the size of the signal search area from 54.3 mm$^2$ in

2003 to 6.4 mm$^2$ in 2004 which is possible due to the monitoring of the stability of the X-ray focus with the pyroelectrical source installed for the 2004 run. The second improvement is a 50% background reduction due to a more complete detector shielding.

IV. CONCLUSION AND OUTLOOK

The combination of a superconducting dipole magnet producing a magnetic field integral of 83 Tesla·m and a focusing X-ray telescope enhancing the signal-to-background ratio by more than 200 has resulted in an axion helioscope of unprecedented sensitivity. The combined result [8] of the 3 detection systems at CAST from data taken in 2003 reduces the previously best upper limit [7] on the axion-photon coupling constant $g_{a\gamma}$ for small axion masses ($m_a < 0.02$ eV) by a factor of 5. The limit obtained in 2003 by the telescope alone ($g_{a\gamma} < 1.23 \times 10^{-10}$ GeV$^{-1}$ at 95 % CL) is already dominating the combined result [8] and with the data taken with the X-ray telescope in 2004, for the first time in a controlled laboratory experiment, the axion sensitivity will supersede the best constraint obtained from astrophysical arguments (compare "CAST prospects" and "globular clusters" in Fig. 3).

Presently, the CAST collaboration is preparing the extension of the solar axion search to higher axion masses (up to about 1 eV) [14]. Filling the conversion volume of the magnet with a low-Z buffer gas (He$^4$ first, then He$^3$) of variable pressure will provide X-ray photons with an effective mass lifting the axion-photon momentum mismatch and restoring particle coherence over the entire length of the magnetic field. An example of the expected sensitivity of a series of overlapping measurements is shown at the high mass end of the line "CAST prospects" in Fig. 3. The low background rates of the telescope achieved in 2004 will allow to scan the accessible mass range almost free of background. The data of this second phase of CAST experiments is expected to probe for the first time into the theoretically motivated range of the axion coupling strength.

ACKNOWLEDGMENT

We thank our colleagues at CAST for their support and cooperation and the CERN management for hospitality.